**A Narrowband Spintronic Terahertz Emitter based on Magnetoelastic Heterostructures**

Shihao Zhuang[1], Peter B. Meisenheimer[2], John Heron[2], Jia-Mian Hu[1*]

[1]Department of Materials Science and Engineering, University of Wisconsin-Madison, Madison, WI, 53705, USA

[2]Department of Materials Science and Engineering, University of Michigan, Ann Arbor, Michigan, 48109, USA

**Abstract**

Narrowband terahertz (THz) radiation is crucial for high-resolution spectral identification, but a narrowband THz source driven by femtosecond (fs) laser has remained scarce. Here, it is computationally predicted that a metal/dielectric/magnetoelastic heterostructure enables converting a fs laser pulse into a multi-cycle THz pulse with a narrow linewidth down to ~1.5 GHz, which is in contrast with the single-cycle, broadband THz pulse from the existing fs-laser-excited emitters. It is shown that such narrowband THz pulse originates from the excitation and long-distance transport of THz spin waves in the magnetoelastic film, which can be enabled by a short strain pulse obtained from fs laser irradiation of the metal film when the thicknesses of the metal and magnetoelastic films both fall into a specific range. These results therefore reveal an approach to achieving optical generation of narrowband THz pulse based on heterostructure design, which also has implications in the design of THz magnonic devices.

[*]E-mail: jhu238@wisc.edu

**Introduction**

The ability to generate terahertz (THz) pulses (pulsed electromagnetic radiation with a frequency of 0.3-3 THz) from visible or near-infrared femtosecond (fs) laser pulses[1,2] is a milestone in THz technology.[3] It has enabled the development of the THz time-domain spectroscopy for spectral identification of a wide variety of chemical species.[4–6] Notably, THz waves can penetrate many optically opaque packaging materials (e.g., paper, fabric, plastics),[7,8] and hence are well-suited for non-destructive evaluation and security imaging.

There are three main types of fs-laser-excited THz emitters, which are based on photoconductive semiconductors (e.g., GaAs, InAs),[9–11] electro-optical crystals (e.g., ZnTe, GaP, $LiNbO_3$),[9,12–14] and ferromagnet(FM)/heavy-metal(HM) bilayer films,[9,15–23] respectively. In the first type, photoexcited electrons accelerate in a bias electric field, forming a time-varying current which generates THz radiation. In the second type, THz radiation arises from optical rectification. The third type is known as a spintronic THz emitter. During operation, the fs laser pulse irradiates and creates a non-equilibrium electron distribution in the ferromagnetic film where the photoexcited majority-spin electrons have much larger velocity than the excited minority-spin electrons. This in turn generates a spin current pulse which transports from the FM into the HM film in a superdiffusive manner[24,25] and has an ultrashort duration of typically <1 picosecond (ps). Through the inverse spin Hall effect in the heavy metal,[26] the ps spin current pulse is converted into a ps charge current pulse, and the emitted THz pulse is considered to arise primarily from the charge current pulse via electric dipole radiation.[15] Although such laser-induced THz spin transport should in principle occur in any metallic FM/HM stacks,[27] its underlying mechanisms await further clarification. For example, it has recently been suggested that the laser-induced THz spin transport in the FM/HM stack shares a common driving force as the laser-induced ultrafast demagnetization in a single FM metal film.[28] It has also been shown that the role of fs-laser-induced lattice strain in the FM metal in its ultrafast spin dynamics cannot be excluded.[29] Nevertheless, it is noteworthy that the THz pulses generated from all three types of emitters are single-cycle, and have broad frequency window due to the short lifetime (at most a few ps),[9] making them suitable for use in broadband THz spectroscopy.[4] However, for applications requiring selective excitation of a particular THz mode such as narrowband spectroscopy[30] and resonant pumping of materials,[31] a narrowband THz pulse would be desirable.

In this article, we computationally design and demonstrate a different type of spintronic THz emitter, which is comprised of a metal/dielectric/magnetoelastic heterostructure and can convert a fs laser pulse into a multi-cycle THz pulse that lasts up to nanoseconds and displays a single peak of narrow linewidth down to ~1.5 GHz, in stark contrast with existing fs-laser-excited emitters which all generate single-cycle, broadband THz pulse lasting at most a few ps. We show that such narrowband THz pulse originates from the excitation and long-distance transport of THz spin waves in the magnetoelastic film, which we predict can be achieved by injecting a short strain pulse obtained from fs-laser-excitation of the metal/dielectric bilayer underneath. Influence of materials selection and parameters on the frequency and energy density of the emitted THz pulse are also discussed. Compared to conventional spintronic THz emitters where the fs laser pulse is pumping the magnetic thin film directly with multiple mechanisms determining the spin dynamics simultaneously, herein the fs laser is fully absorbed by the metal transducer for generating ultrashort strain pulse.[32–34] By employing a thick dielectric layer as a heat sink, the spin dynamics in the magnetoelastic film would be solely modulated by acoustic strain without being affected by laser-induced heating. As a result, the features of the THz pulse emitted from the present acoustically mediated spintronic THz emitter are entirely different.

**Results**

An Al (10nm)/(001)MgO/(001)Fe (900nm) epitaxial magnetoelastic heterostructure is used as an example. The parameters of the fs laser (wavelength: 800 nm; duration: 20 fs; absorbed pulse energy: 13 J m$^{-2}$) are the same as those in a conventional spintronic THz emitter based on Glass(substrate)/Fe(10 nm)/Au(2nm) heterostructure.[15] Al is a commonly used metal transducer for converting a fs laser pulse into a ps acoustic pulse[32–34] because of its large thermal expansion and small absorption length to the near-infrared laser. The dielectric (001) MgO substrate confines laser-excited hot electrons within the Al film for quickly raising the lattice temperature of Al, shields the Fe film from the heat deposited to the Al, and enables the epitaxial growth of (001) Fe film.[35]

Let us now consider that the fs laser pulse uniformly irradiates the entire bottom surface of the metallic Al, as schematically shown in Fig. 1A. Since the Al film exhibits an absorption length of ~7.86 nm to the 800-nm-wavelength laser,[36] a 10-nm-thick Al film should absorb the majority of the incident photons, leading to a rapid rise of the electron temperature in the Al film via photoelectric excitation and then its lattice temperature via electron-phonon coupling. As a result, a short longitudinal strain pulse $\varepsilon_{zz}(z,t)$ is injected from the Al into the MgO due to the lattice and electronic thermal stress. Figure 1B shows the spatial profiles of the $\varepsilon_{zz}$ along the thickness direction of the Al/MgO bilayer at 10, 20, and 30 ps after irradiation. Such strain profiles are simulated by coupling the two-temperature model,[32,37] elastodynamics, and heat transport (see Methods). Figure 1C further shows the spatiotemporal profile of the $\varepsilon_{zz}$ inside the Fe film, which is obtained by phase-field simulations that consider the magnetoelastic backaction of magnetization dynamics on elastic waves (see Methods). This is necessary for realizing an accurate modeling of both the spin and lattice dynamics in magnetic materials.[38–40] As shown in Fig. 1C, the fs-laser-induced strain pulse $\varepsilon_{zz}(z,t)$ propagates into the Fe film from its bottom surface ($z=0$) at $t=0$ ps, reaches the free top surface at $t\sim180$ ps, propagates backwards after reflection, and reaches the bottom surface again at $t\sim360$ ps. After that, most of the strain pulse leaves the Fe film. Accordingly, the elastically excited spin wave $\Delta m_y(z,t)$ is stronger within the time span of $t=0$-360 ps, as shown in Figure 1D. In particular, the fastest-propagating spin wave arrives at the top Fe surface at $t\sim90$ ps, about two times faster than the strain pulse.

**Mechanisms of spin wave excitation**

Figure 2A shows the spin wave dispersion relation $f(k)$ obtained by performing 2D Fourier transform of the $\Delta m_y(z,t)$ data over the entire Fe film thickness for the full time span of $t=0$-720 ps, with two enlarged plots near the two crossing points ($X_1$ and $X_2$) displayed on the side. As shown, a bright spot appears at 1.04 THz, indicating the presence of a THz spin wave mode that carries a significant portion of energy in the entire wave packet. The linear portion of the $f(k)$ well fits the formula $f=v_s k$ (see dotted straight lines), where $v_s$ =5357 m/s is the longitudinal sound velocity in Fe. The spin wave components displaying such linear $f(k)$ relation travel at the same speed as the strain pulse, indicating a pure magnetoelastic excitation mechanism. This interpretation is supported by a control simulation performed under zero exchange coupling coefficient $A_{ex}$ in which case the simulated $f(k)$ displays a linear relation but only with very weak amplitude at frequencies near 1 THz, as shown in Fig. 2B. This finding is consistent with the frequency spectrum of the injected strain pulse $\varepsilon_{zz}(z,t)$ shown in the inset of Fig. 2B, where the amplitude of the elastic wave component at near 1 THz is found to be weak as well. Comparing Figs. 2A and 2B, it can be concluded that the desirable strong 1.04 THz spin wave mode cannot be excited via magnetoelastic coupling alone.

The parabolic curve of the $f(k)$ in Fig. 2A well fits the formula $f=f_0+Dk^2$ (see dotted curve),[41,42] where the coefficient $D = \frac{4\pi A_{ex}\gamma}{\mu_0 M_s}$; $f_0$ is the ferromagnetic resonance (FMR) frequency that can be calculated via Smit-Beljers formula.[43] For (001)Fe, $f_0 = \frac{\gamma}{2\pi}\sqrt{(K_1/2\mu_0)-(2K_1^2/\mu_0^2 M_s^2)}$ = 4.73 GHz. Here, $\gamma$ is the gyromagnetic ratio; $\mu_0$ is the vacuum permeability; $M_s$ is the saturation magnetization; $K_1$ is the magnetocrystalline anisotropy coefficient. Therefore, spin wave components displaying such parabolic $f(k)$ relation are excited by exchange coupling. Among all the exchange spin wave components of different wavenumbers and frequencies being excited, only those having the same wavenumber and frequency as those of the injected strain pulse can propagate with gradually increasing amplitude due to the in-phase construction, hence they can propagate for a relatively long distance against damping-induced attenuation. The phase velocities of these exchange spin wave components ($v_{p,m}=f/k$) are equal to the phase velocity of the injected strain pulse, which is the same as its group velocity (i.e., the sound velocity $v_s$) since the injected strain pulse has a linear $f(k)$ relation. In this regard, the wavenumber $k^*$ and frequency $f^*$ of these exchange spin waves can be determined analytically by letting $v_{p,m}=f^*/k^*=f_0/k^*+Dk^*=v_s$, with two solutions $k^*=\left(v_s\pm\sqrt{v_s^2-4Df_0}\right)/2D$. Plugging in the relevant material parameters of (001) Fe (see Supplementary Table S1), one has $k_1^*$ = 1.95×10$^8$ m$^{-1}$ and $k_2^*$ = 8.87×10$^5$ m$^{-1}$, corresponding to a high-frequency exchange spin wave $f_1^*$ =1.046THz and a low-frequency exchange spin wave $f_2^*$=4.75GHz, respectively. These two analytically calculated frequencies and wavenumbers agree well with the values at the two crossing points of the simulated spin wave dispersion relation, as shown by the two enlarged plots in Fig. 2A. Furthermore, the group velocities of the low-frequency and high-frequency exchange spin wave can be analytically calculated via $v_g=(\partial f/\partial k)_{k=k^*}=2Dk^*$, with $v_g$ =48 m/s and 10665.6 m/s, respectively. Therefore, the 1.04 THz exchange spin wave has a group velocity about twice that of the injected strain pulse and hence would propagate ahead.

For further clarifying the mechanism of spin wave excitation, the real-space profiles of the spin wave and strain pulse in the Fe film are shown in Fig. 2C. After $t$=0 ps, the moment strain pulse propagates into the Fe film from its bottom surface ($z$=0), local magnetization vector therein will start to precess almost immediately due to magnetoelastic coupling. Through exchange coupling, the nearest magnetization vector will then precess, and so forth for the second nearest magnetization vector, leading to the excitation of an exchange spin wave. As the strain pulse travels across the Fe, it constantly excites new exchange spin waves of different wavenumbers and frequencies propagating along the Fe film thickness direction. The location of the faster-propagating 1.04 THz exchange spin wave is indicated in the overall waveform, which clearly shows that its amplitude increases over time due to the in-phase construction. The wavelength of the 1.04 THz exchange spin wave is 5.2 nm as indicated in Fig. 2C. It agrees well with the analytically calculated wavenumber $k_1^* = 1.95 \times 10^8$ m$^{-1}$, which gives wavelength $1/k_1^*$ of 5.13 nm. By drawing the trajectories of the frontmost positions of the 1.04 THz exchange spin wave and strain pulse (dashed lines in Fig. 2C), it can be estimated from their slope that the group velocity of the former is about twice larger, consistent with the analytical calculation. It is worth adding that the above-discussed mechanism of exciting propagating THz exchange spin wave by a short strain pulse neither involves nor relies on the formation of hybridized magnon-phonon mode (or magnon polarons), which would lead to the presence of two anticrossing regions in the spin wave dispersion relation[41,44,45] rather than the two crossing points herein. In fact, according to our control simulation where magnetization-related terms are dropped when solving the elastodynamic equations (Eqs. 1-3) and hence the formation of magnon polarons is mathematically impossible, the desirable propagating THz exchange spin wave can still be excited and the obtained spin wave dispersion relation remains virtually the same (see Supplementary Fig. S1).

We note that the Al film needs to be sufficiently thin to ensure that the amplitude of such 1.04 THz exchange spin wave is appreciable. Specifically, a thinner Al film allows for generating a faster-rising and larger strain pulse, which can induce a larger-amplitude, shorter-wavelength spin wave via magnetoelastic coupling. This will further yield a larger exchange coupling field, thereby enhancing the amplitude of the exchange spin wave. As shown in Supplementary Figure S2, the amplitude of the faster-propagating exchange spin wave is barely appreciable in the case of 100-nm-thick Al, but increases significantly as the Al film thickness decreases.

We also note that the Fe film needs to be sufficiently thick to ensure that the faster-propagating 1.04 THz exchange spin wave has a long duration (denoted as $t_{ESW}$) such that it can constitutes a major portion in the entire spin wave packet. Specifically, this requires $t_{ESW} > t_{MSW}$; here $t_{MSW}$ is the duration of the lower-frequency magnetoelastic spin wave, which is the time required for completing the reflection of the strain pulse from the Fe top surface. For the strain pulse shown in Fig. 1B, $t_{MSW}$ ~ 16 ps. The duration of the 1.04 THz exchange spin wave can be estimated as $t_{ESW}$~$d/v_g$ (see Supplementary Figure S3), which is the time span from the moment it arrives at the Fe top surface to the moment the strain pulse arrives. Accordingly, the Fe film thickness needs to be thicker than ~170 nm (=$v_g \times t_{MSW}$).

**THz radiation from THz spin waves**

Such elastically excited THz exchange spin waves can generate strong THz emission via magnetic dipole radiation. Our discussion will be based on the near-field electric-field component of the emitted electromagnetic wave because THz radiation is usually detected by near-field electro-optical sampling. Figure 3A shows the evolution of the electric field component of the radiation $E_x(t)$ above the Fe top surface, which is proportional to the spatial gradient $\partial_z m_y(z,t)$. Three distinct wave packets are shown over the time span of $t$=0-360 ps. The first wave packet of $E_x(t)$ starts from the moment the strain pulse enters the Fe bottom surface ($t$=0 ps) to the moment its entire waveform propagates into the Fe film ($t$=16 ps). This $E_x(t)$ packet results primarily from the excitation of low-frequency magnetoelastic spin wave near the Fe bottom surface and displays a broadband feature in the frequency range of 0-0.56 THz with a peak frequency of 0.14 THz (see Supplementary Figure S4). The second wave packet of $E_x(t)$ starts from the moment the faster-propagating 1.04 THz exchange spin wave arrives at the Fe top surface ($t$=84 ps) to the moment the strain pulse $\varepsilon_{zz}(z,t)$ is completely reflected from the Fe top surface ($t$=184 ps). As shown in Fig. 3A, the second wave packet of $E_x(t)$ comprises a monochromatic high-frequency phase (84-168 ps) and a non-monochromatic low-frequency phase (168-184 ps). In the first phase, the strain pulse has not yet arrived at the Fe top surface, the radiation arises from a THz standing spin wave that forms due to the interference between the incident and reflected THz exchange spin waves. The formation of such a standing spin wave can be seen from the spatiotemporal profile of the $\Delta m_y(z,t)$ near the Fe top surface ($z$=870-900 nm) within $t$=84-104 ps, as shown in Fig. 3B. As the magnitude of the standing spin wave increases, the magnitude of $E_x(t)$ increases accordingly. The wavelength of the standing spin wave is 5.2 nm, corresponding to a wavenumber of $k^* = 1.92 \times 10^8$ m$^{-1}$ which well agrees with the value of $1.95 \times 10^8$ m$^{-1}$ calculated analytically via the formula $k^* = \left(v_s + \sqrt{v_s^2 - 4Df_0}\right)/2D$ derived earlier. By contrast, the standing spin waves excited by short strain pulse in relatively thin (<20 nm) magnetoelastic films[46] exist across the entire film thickness, and have discrete wavenumbers that are inversely proportional to the film thickness $d$ with $k=n\pi/d$ ($n$=1,2,3…). In the second phase (168-184 ps), the flipped waveform of the strain pulse after the reflection from the top Fe surface causes an abrupt flipping of the waveform of spin wave $\Delta m_y(z,t)$, leading to a spike in the $E_x(t)$ at $t$=171 ps. Once the backward-propagating spin wave $\Delta m_y(z,t)$ arrives at the Fe bottom surface ($t$=252 ps) and gets reflected again, standing spin wave with a frequency of 1.04 THz would form near the Fe bottom surface (see Supplementary Figure S5). However, the 900-nm-thick metallic Fe will significantly absorb the electromagnetic wave emitted from the Fe bottom surface, leading to relatively small $E_x(t)$ at above the Fe top surface, as shown in the third wave packet of $E_x(t)$ in Fig. 3A. In particular, since metals absorb higher-frequency electromagnetic waves more significantly than lower-frequency ones, the high-frequency $E_x(t)$ is barely appreciable. As a result, only the second wave packet of $E_x(t)$ displays a strong THz frequency component.

Figure 3C shows the frequency spectrum, $|E_x(f)|$, of the second wave packet of the $E_x(t)$. A single peak with a narrow linewidth of ~0.02 THz appears at 1.04 THz, which corresponds to the 1.04 THz standing spin wave. The broadband feature in the lower-frequency regime is related to the lower-frequency magnetoelastic spin wave. As shown by the energy spectral density in Figure 3D, which is proportional to $|E_x(f)|^2$, the majority of the energy is concentrated within the peak of 1.04 THz. The energy density of the 1.04 THz emission can be quantified as $e_{THz} = \int_{t_{start}}^{t_{end}} |E_x(t)|^2/\eta_0 dt$, where $\eta_0$=377 Ω is the impedance of free space, $t_{start}$=84 ps and $t_{end}$=168 ps marks the start and end of the high-frequency THz wave. The calculated $e_{THz}$ is 43.9 nJ m$^{-2}$, which is about 74% of the total energy density contained in the second wave packet. Given that the laser pulsed energy absorbed by the Al film is 13 J/m$^2$, this gives an energy conversion efficiency of ~3.4×10$^{-9}$ from the Al(10 nm)/MgO/Fe(900 nm) heterostructure, which is two orders of magnitude larger than that from a conventional spintronic THz emitter based on glass/Fe(10 nm)/Au (2 nm) heterostructure under the same absorbed laser energy.[15] This is however mainly because the THz emission in the latter case lasts for only ~2 ps (see Supplementary Fig. S6), where a transient THz pulse is desirable for applications like broadband spectroscopy. Next, we will discuss the principles for further enhancing the $e_{THz}$ and hence the energy conversion efficiency of the present narrowband spintronic THz emitter.

**Enhancing the THz emission by heterostructure design**

Having demonstrated fs-laser-excited narrowband THz emission in the Al/MgO/Fe heterostructure, we further explore four other metal/dielectric/magnetoelastic heterostructures including Al/MgO/Fe$_{60}$Co$_{40}$, Al/MgO/Fe$_{80}$Ga$_{20}$, Al/MgO/Co$_{20}$Fe$_{60}$B$_{20}$, and Al/Gd$_3$Ga$_5$O$_{12}$/Y$_3$Fe$_5$O$_{12}$. Both the dielectric substrate and the magnetoelastic film are (001)-oriented. All material parameters used for computations are experimentally measured values, which are summarized in Supplementary Table S1 and Table S2. These magnetoelastic films cover a broad range of the magnitude of magnetoelastic coupling coefficient $|B_1|$ from 0.3 MJ m$^{-3}$ in Y$_3$Fe$_5$O$_{12}$ (YIG) to 29.3 MJ m$^{-3}$ in Fe$_{60}$Co$_{40}$, as well as the damping parameter $\alpha$ from $8\times10^{-5}$ in YIG to 0.017 in Fe$_{80}$Ga$_{20}$. A lower $\alpha$ enhances both the duration and amplitude of the THz exchange spin wave and hence the THz emission (see Supplementary Figure S7). A larger $|B_1|$, on one hand, yields a larger magnetoelastic field that increases the spin wave amplitude and hence the emitted electric field, and on the other hand, leads to larger secondary elastic waves via magnetoelastic backaction that suppress the rise of strain and hence the excitation of THz exchange spin wave (see Supplementary Video 1). As a result, a moderately large $|B_1|$ is preferred for enhancing the THz emission (see Supplementary Figure S8).

Starting from the same absorbed laser fluence (13 J m$^{-2}$) and laser pulse duration (20 fs), the frequency $f^*$ and energy density $e_{\text{THz}}$ of the THz emission for all five magnetoelastic heterostructures are shown in Figure 4A. Typical profiles of $E_x(t)$ and corresponding frequency spectra are shown in Supplementary Figure S9. For each heterostructure, varying the thickness of the magnetoelastic film allows for tuning both the duration and magnitude of the THz electric field and therefore the $e_{\text{THz}}$. However, the frequency $f^*$ for each heterostructure, which agree well with the value calculated analytically via $f^*=v_s\left(v_s+\sqrt{v_s^2-4Df_0}\right)/2D \approx v_s^2/D$, remains unchanged since the relevant material parameters are assumed to be independent of the film thickness. In particular, the $e_{\text{THz}}$ emitted from the YIG based heterostructures can reach the order of $10^{-4}$ J m$^{-2}$. This corresponds to an energy conversion efficiency of $\sim7.69\times10^{-6}$, which is about three orders of magnitude larger than the other four heterostructures. Given that $e_{\text{THz}}$ is proportional to both the duration of the emitted THz electric field and the square of its magnitude, this finding can be analyzed from the following two aspects. First, the magnetic damping coefficient of the (001) YIG is two-to-three orders of magnitude smaller than the others (see Table S1), leading to a ns-long duration of the THz exchange spin wave and hence the THz radiation. By contrast, the duration in the remaining four cases varies from 50 ps to 180 ps. Second, it is found that the amplitude of the THz electric field emitted from the insulating YIG is larger than the other four conducting magnets investigated (Fe, $Fe_{60}Co_{40}$, $Fe_{80}Ga_{20}$, $Co_{20}Fe_{60}B_{20}$). This is because a large portion of the emitted electromagnetic wave would be absorbed by the relatively thick (thickness: 200 nm-2000 nm) magnetic conductors themselves via eddy current loss.

As an example, Figure 4B shows the profile of $E_x(t)$ from a 20-µm-thick YIG film, where the high-frequency ($f^*$=1.01 THz) phase of the $E_x(t)$ emitted has a duration of ~1.4 ns (from 1.386 ns to 2.773 ns) and an average amplitude of 1.73 kV/m. This duration is one order of magnitude larger than the 84 ps in the case of 900-nm-thick Fe (c.f., Fig. 3A), and the average amplitude is also one order of magnitude larger. However, if not considering the eddy current loss in the Fe film, the calculated high-frequency $E_x(t)$ from the Al/MgO/Fe (900 nm) heterostructure would be one order of magnitude larger than the YIG-based one (see Supplementary Fig. S10). Figure 4C shows the frequency spectrum of the second wave packet of $E_x(t)$ in Fig. 4B. Since the time-domain waveform is dominated by the 1.4-ns-duration 1.01-THz electric field (the duration of the lower-frequency phase is only 0.016 ns), a single peak with an ultra-narrow linewidth of ~1.5 GHz is shown in the frequency spectrum. In addition to THz oscillation, $E_x(t)$ also displays a low-frequency oscillation with a frequency of ~0.5 GHz. This value is consistent with the ferromagnetic resonance (FMR) frequency calculated analytically via the Smit-Beljers formula,[43] which describes the precession of local magnetization driven by local magnetocrystalline anisotropy field and magnetic dipolar coupling field. Detailed discussion can be found in Supplementary Note 1.

Figure 5 shows the frequency $f^*$ and linewidths of the THz pulses from the present narrowband spintronic THz emitter and the existing fs-laser-excited THz emitters. Data points for the latter were collected from refs. [9,10,12,13,15,20,47–51], and their corresponding numerical values and specific materials are listed in supplementary Table S4. As shown in Fig. 5, the frequencies of the five narrowband spintronic THz emitters vary from 0.7 to 1.1 THz, which cover a smaller frequency window than the three existing technologies. Although it is possible to tune the frequency by materials engineering based on the simple formula $f^* \approx v_s^2/D = v_s^2 \frac{\mu_0 M_s}{4\pi A_{ex}\gamma}$ discovered in this work, it may still be challenging to elevate the frequency to be >2 THz using the magnetoelastic materials investigated here. A more notable observation is the one-to-three orders of magnitude narrower linewidth, which is mainly due to the long-duration (up to ns), multi-cycle nature of the emitted THz pulse. In this regard, the present narrowband spintronic THz emitter well complements the existing broadband THz emitters, and may find use in application scenarios that require narrowband THz pulses such as narrowband spectroscopy. Importantly, the computational results in this work establish the principles of tuning the frequency, linewidth, and the energy density of the THz pulse from the present new-concept spintronic THz emitter by heterostructure design, which provide guidance for subsequent experimental validation.

**Conclusions**

In summary, we have computationally demonstrated a spintronic THz emitter which consists of a metal/dielectric/magnetoelastic heterostructure and can convert a fs laser pulse into a multi-cycle, narrowband THz pulse via the acoustic excitation of THz spin waves, in contrast with single-cycle, broadband THz pulses from the existing fs-laser-excited emitters.[9] We have also shown that the frequency, linewidth, and energy density of the THz pulse can be tuned by heterostructure design. Beyond THz technology, our simulations also reveal an approach to achieving long-distance transport of a THz exchange spin wave by constantly exciting new THz exchange spin waves of the same frequency and phase using a travelling strain pulse (see Fig. 2C). This in effect extends the decay length of THz exchange spin wave to that of elastic waves, which can be up to millimeters or longer. For example, in the case of a 20-µm-thick YIG film (Fig. 4B), such THz exchange spin wave can propagate 20 µm and so forth. The ability to achieve millimeter-long transport of THz exchange spin waves, which by themselves can only propagate a few nanometers in 3d ferromagnets before decaying into other lower-energy modes,[52] offers new opportunities for designing THz magnonic devices.[53]

**Methods**

**Modeling fs-laser-induced excitation of ultrafast strain pulse**

The process of generating ultrafast (typically ps scale) strain pulse in the metal/oxide bilayer by fs laser pulse is modelled by coupling the classical two-temperature model (which describes the electron-phonon coupling),[32,37] the heat transport equation, and the elastodynamic equation incorporating the lattice and electronic thermal stress. The results are shown in Fig. 1B. Details can be found in Supplementary Note 2. The high numerical accuracy of our codes was demonstrated by benchmarking against the finite-element-method bases solvers in commercial software COMSOL Multiphysics using specifically designed test problems (Supplementary Note 3). Our calculations are performed by considering the irradiation from a single fs laser pulse (wavelength: 800 nm; duration: 20 fs; absorbed pulse energy: 13 J m$^{-2}$). The laser-induced ps strain pulse travels across the dielectric layer and propagates into the overlaying magnetoelastic film, where its effects on the local magnetization dynamics are simulated using a dynamical phase-field model described below. Since the dielectric layer is thick enough to shield the magnetoelastic film from the heat deposited to the metal transducer, the effects of laser-induced heating on the spin dynamics are not considered.

**Dynamical Phase-field model coupling micromagnetics with elastodynamics**

Standard micromagnetic models can consider the effects of time-varying elastic wave on the magnetization dynamics by modulating the magnetoelastic anisotropy, but the back-action of local magnetization dynamics on the elastic wave through the magnetoelastic feedback[54] is omitted. An accurate modelling of the mutually coupled spin and elastic waves requires solving the equations of motion for local magnetization and local mechanical displacement in a coupled fashion,[38–40] which is particularly important for material with strong magnetoelastic coupling. In this work, an in-house dynamical phase-field model that considers the coupled spin and lattice dynamics in a heterostructure with magnetic and elastic inhomogeneity is developed. In this model, the evolution of local mechanical displacement **u** is described by elastodynamic equation incorporating the magnetostrictive stress, given by,

$$\rho \frac{\partial^2 u_x}{\partial t^2} = c_{44} \frac{\partial^2 u_x}{\partial z^2} + B_2 \frac{\partial (m_x m_z)}{\partial z}, \tag{1}$$

$$\rho \frac{\partial^2 u_y}{\partial t^2} = c_{44} \frac{\partial^2 u_y}{\partial z^2} + B_2 \frac{\partial (m_y m_z)}{\partial z}, \tag{2}$$

$$\rho \frac{\partial^2 u_z}{\partial t^2} = c_{11} \frac{\partial^2 u_z}{\partial z^2} + B_1 \frac{\partial (m_z^2)}{\partial z}, \tag{3}$$

where $\rho$ is the mass density; $c_{11}$, $c_{12}$ and $c_{44}$ are the elastic stiffness coefficients. $\rho$ and **c** are different in different layers of the metal/dielectric/magnetoelastic heterostructure. $B_1 = -1.5\lambda_{100}(c_{11}^M - c_{12}^M)$ and $B_2 = -3\lambda_{111} c_{44}^M$ are the magnetoelastic coupling coefficients of the magnetoelastic film ($\lambda_{100}$ and $\lambda_{111}$ are its magnetostrictive coefficients); **m**=**M**/$M_s$ is the normalized local magnetization vector. From Equation (3), it can be seen that longitudinal elastic wave $\varepsilon_{zz}(z,t)$ in the magnetic layer should include a magnetoelastic feedback from time-varying magnetization.[40,55] Such secondary elastic waves have significant influence on the exchange spin wave excitation as shown in Supplementary Figure S8 and Supplementary Video 1. The evolution of local magnetization $\mathbf{m}(z,t)$ is governed by the Landau-Lifshitz-Gilbert (LLG) equation,

$$\frac{\partial \mathbf{m}}{\partial t} = -\frac{\gamma}{1+\alpha^2}\mathbf{m}\times\mathbf{H}_{\text{eff}} - \frac{\alpha\gamma}{1+\alpha^2}\mathbf{m}\times(\mathbf{m}\times\mathbf{H}_{\text{eff}}). \qquad (4)$$

The total effective magnetic field $\mathbf{H}_{\text{eff}}$ is contributed by the magnetocrystalline anisotropy field $\mathbf{H}^{\text{anis}}$, magnetic exchange coupling field $\mathbf{H}^{\text{exch}} = \frac{2A_{\text{ex}}}{\mu_0 M_s}\nabla^2\mathbf{m}$, magnetoelastic anisotropy field $\mathbf{H}^{\text{mel}}$, magnetic dipolar coupling field $\mathbf{H}^{\text{dip}}$, and the external magnetic field $\mathbf{H}^{\text{ext}}$. For magnetic materials with a cubic high-temperature parent phase, one has ($i = x, y, z,$ and $j \neq i, k \neq i, j$),

$$H_i^{\text{anis}} = -\frac{2}{\mu_0 M_s}\left[K_1\left(m_j^2+m_k^2\right)+K_2 m_j^2 m_k^2\right]m_i, \qquad (5)$$

$$H_i^{\text{mel}} = -\frac{2}{\mu_0 M_s}\left[B_1 m_i \varepsilon_{ii} + B_2\left(m_j \varepsilon_{ij} + m_k \varepsilon_{ik}\right)\right], \qquad (6)$$

where $\mu_0$ is vacuum permeability; $K_1$ and $K_2$ are magnetocrystalline anisotropy coefficients; strain $\varepsilon_{ij} = \frac{1}{2}\left(\frac{\partial u_i}{\partial j}+\frac{\partial u_j}{\partial i}\right)$. Since the fs-laser-induced ultrafast acoustic pulse in the present Al(polycrystal)/dielectric(001)/magnetoelastic(001) heterostructure only contains the component $\varepsilon_{zz}$, the magnetoelastic field in the magnetoelastic film is given by $(H_x^{\text{mel}}, H_y^{\text{mel}}, H_z^{\text{mel}}) = (0, 0, -\frac{2B_1}{\mu_0 M_s}m_z\varepsilon_{zz})$. Therefore, the initial magnetization state must have a nonzero out-of-plane component ($m_z \neq 0$), otherwise the $\varepsilon_{zz}(t)$ cannot interact with the local magnetization. One must also ensure that the initial magnetization has a nonzero in-plane component ($m_x \neq 0$ or $m_y \neq 0$), otherwise the torque $\mathbf{m}\times\mathbf{H}^{\text{mel}} = (0,0,m_z)\times(0,0,H_z^{\text{mel}})$ would be zero and hence $\mathbf{m}$ would not evolve. Thus, we introduce a bias magnetic field along the $z$-direction ($\mathbf{H}^{\text{ext}} = H_z^{\text{ext}}$) to obtain a titled, spatially uniform initial magnetization $\mathbf{m}$ in an otherwise in-plane magnetized film. The tilting angle of the initial $\mathbf{m}$ is 45° off the film plane, in which case the torque from the magnetoelastic field $\mathbf{m}\times\mathbf{H}^{\text{mel}} = \frac{2B_1}{\mu_0 M_s}\left(-m_y m_z \varepsilon_{zz}, m_x m_z \varepsilon_{zz}, 0\right)$ is maximized. Moreover, for an infinitely large $xy$ plane within which the magnetization $\mathbf{m}$ is spatially uniform, the dipolar coupling field is calculated as $\mathbf{H}^{\text{dip}} = (0, 0, -M_s m_z)$, which is numerically validated using the open-source software MuMax3 (Supplementary Note 4). The material parameters for all six heterostructures are listed in Supplementary Table S1 and Table S2.

When numerically solving Equations (1-6), the metal/dielectric/magnetoelastic heterostructure is discretized into one-dimensional (1D) computational cells along $z$ direction, with cell size $\Delta z = 0.2$ nm. The dielectric substrate needs to be thick enough to ensure that (i) the temperature increase is not transported into the magnetoelastic film; and (ii) the portion of the acoustic pulse reflected back from the dielectric/magnetoelastic interface will not enter the magnetoelastic layer before the major portion of the acoustic pulse completely leaves the magnetoelastic film. Central finite difference is used for calculating spatial derivatives. Equations (1-6) are coupled with each other and solved simultaneously, classical Runge-Kutta method is used for time-marching with a real-time step $\Delta t = 0.2$ fs. The same results are obtained from simulations with $\Delta t = 0.1$ fs. Appropriate magnetic, and mechanical boundary conditions are implemented on all surfaces and interfaces. These include (i) the free magnetic boundary condition $\partial \mathbf{m}/\partial \mathbf{n}=0$ at the top and bottom surface of the magnetic layer, which leads to the reflection of the excited spin waves between its two surfaces; (ii) displacement continuity and stress continuity, $\mathbf{u}^A=\mathbf{u}^B$, $\boldsymbol{\sigma}^A=\boldsymbol{\sigma}^B$; superscripts 'A,B' denote different materials. In particular, by setting the stress in the air to be zero, the continuity boundary condition is converted to a stress-free boundary condition $\boldsymbol{\sigma}^A=0$ (precisely, $\sigma_{iz}^A=0$ at the outermost layer, $i=x,y,z$) which leads to the reflection of elastic wave from the top surface of the magnetic layer. The laser-induced ultrafast acoustic pulse is fed as the initial displacement distribution to the phase-field model.

**Calculating emitted electric field from magnetic dipole radiation**

The electric field component $E_x(t)$ of the magnetic dipole radiation is obtained by an inverse Fourier transformation of its frequency domain component $E_x(\omega)$, which is obtained via numerically solving the plane wave equation,

$$\frac{\partial^2 E_x(z,\omega)}{\partial z^2}+\omega^2\mu_0\varepsilon_0\varepsilon_r(\omega)E_x(z,\omega)=i\mu_0\omega\partial_z M_y(z,\omega), \tag{7}$$

where $\varepsilon_0$ and $\mu_0$ are vacuum permittivity and permeability, $\varepsilon_r(\omega)$ is frequency-dependent relative permittivity containing both the real and imaginary parts, with $\varepsilon_r(\omega)= \varepsilon_r^{'}(\omega)+i\varepsilon_r^{''}(\omega)$. $\partial_z M_y(z,\omega)$ is obtained by Fourier transformation of $\partial_z M_y(z,t)$ across the entire magnetoelastic film and over the entire time span, which is extracted from the solution of LLG equation (Equation (4)). $E_x(\omega)$ is taken as the calculated $E_x(z, \omega)$ at the position 2 nm above the top surface of the magnetoelastic film (namely, the detection spot, see Fig. 1A). Although the conducting metal layer of the metal/dielectric/magnetoelastic heterostructure will reflect a portion of the emitted electromagnetic wave, the reflected electromagnetic wave would be reflected again or absorbed by the relatively thick (≥200 nm) magnetoelastic film if the latter is also a conductor (Fe, $Fe_{60}Co_{40}$, $Fe_{80}Ga_{20}$, $Co_{20}Fe_{60}B_{20}$). The calculated free-space electric field at the detection spot, which locates on the other side of the magnetoelastic film, would remain unaffected according to our control simulation (see Supplementary Fig. S11). Only when the magnetoelastic film is an insulator (YIG), the reflected electromagnetic wave could penetrate the magnetoelastic film and interfere with the originally emitted electromagnetic wave at the detection spot. Depending on the thickness of the underlying dielectric layer (GGG), which separates the metal and the magnetoelastic insulator and hence affects the nature of the interference, the free-space electric field at the detection spot could have a similar, smaller, or larger (up to twice larger) amplitude than the case without the reflection from the metal layer (see Supplementary Fig. S12). For ensuring a relatively fair comparison between heterostructures based on magnetic insulators and conductors, the reflection of the electromagnetic wave by the metal is omitted by utilizing a simple air/magnet/air tri-layer for calculation. In practice, the THz electric field detected at above the YIG free surface can be further enhanced by tuning the thickness of the GGG underlayer to enable constructive interference between the reflected and originally emitted THz electromagnetic waves.

For magnetic conductors, the real and imaginary parts of the $\varepsilon_r(\omega)$ are expressed as $\varepsilon_r'(\omega)=1-\omega_p^2\tau^2/(1+\omega^2\tau^2)$ and $\varepsilon_r''(\omega)=\omega_p^2\tau/[\omega(1+\omega^2\tau^2)]$, respectively, based on the Drude model.[56] Here, $\omega_p$ and $\tau$ denote the plasma frequency and electron relaxation time of the metal, respectively. This allows us to consider both the absorption and reflection of the electromagnetic wave by the magnetic conductors. Specifically, the $\varepsilon_r''(\omega)$ is related to the generation of eddy current $J_e = \omega\varepsilon_0\varepsilon_r''(\omega)E_x(z,\omega)$, which leads to the absorption and reflection of emitted electromagnetic wave. Note that the relative permittivity of alloy is assumed to be composition-weighted average of the $\varepsilon_r(\omega)$ for metallic components.[57] For example, relative permittivity of $Co_{20}Fe_{60}B_{20}$, $\varepsilon_r^{CFB}(\omega) = 0.2\ \varepsilon_r^{Co}(\omega) + 0.6\ \varepsilon_r^{Fe}(\omega) + 0.2\varepsilon_r^{B}(\omega)$, where $\varepsilon_r^{Co}(\omega)$ and $\varepsilon_r^{Fe}(\omega)$ are calculated as described above, $\varepsilon_r^{B}(\omega)$ of insulator boron is assumed to be 1. The $\omega_p$ and $\tau$ for all metallic components used in this work are listed in Supplementary Table S3. For magnetic insulators such as YIG, we set that $\varepsilon_r(\omega)=1$ ($\varepsilon_r'(\omega)=1$, $\varepsilon_r''(\omega)=0$) such that the emitted radiation is neither absorbed nor reflected. The high numerical accuracy of our in-house solver for Equation (7) was benchmarked against the finite-element-method bases solvers in commercial software COMSOL Multiphysics using specifically designed test problems (Supplementary Note 5).

**Supporting Information**

Figure S1. Frequency-wavenumber relationship $f(k)$ of spin wave calculated without considering the backaction of acoustically excited spin waves on local strain.

Figure S2. Influence of Al film thickness on the excitation of the THz exchange spin wave

Figure S3. Influence of the Fe film thickness on the excitation of the THz exchange spin wave

Figure S4 The first wave packet of the $E_x(t)$ emitted from the Al(10nm)/(001)MgO/Fe(900nm) heterostructure calculated under zero exchange coupling coefficient

Figure S5. Additional data showing the formation of standing spin wave near the bottom surface of the (001)Fe film

Figure S6. Comparison of the $E_x(t)$ emitted from the conventional broadband spintronic THz emitter (STE) and the present narrowband STE

Figure S7. Influence of magnetic damping coefficient on THz spin wave excitation and resultant THz emission

Figure S8. Influence of magnetoelastic coupling coefficient on THz spin wave excitation and resultant THz emission

Figure S9. Typical profiles of $E_x(t)$ and corresponding frequency spectra from other magnetoelastic heterostructures

Figure S10. Influence of the absorption of electromagnetic wave by magnetic conductor on the amplitude of emitted electric-field component

Figure S11. Influence of the electromagnetic wave reflected by Al on the electric-field profile for magnetic-conductor based emitter

Figure S12. Influence of the electromagnetic wave reflected by Al on the electric-field profile for magnetic-insulator based emitter

Figure S13. Experimentally measured in-plane magnetic hysteresis loops for (001) $Fe_{80}Ga_{20}$ film grown on (001) MgO substrate, through which we extracted the magnetic parameter for computation.

Table S1-3 Summary of material parameters used for computation.

Table S4 Numerical values for the data points in Figure 5 and relevant metadata

Supplementary Note 1. Low-frequency oscillation in the emitted $E_x(t)$

Supplementary Note 2. Modeling fs-laser-induced ultrafast acoustic pulse

Supplementary Note 3. Numerical validation of our in-house acoustic pulse excitation model

Supplementary Note 4. Numerical validation of the analytical expression of the dipolar coupling field

Supplementary Note 5. Calculating emitted electric field from magnetic dipole radiation and numerical validation

**Acknowledgements**

J.-M.H. acknowledges support from the Accelerator Program from the Wisconsin Alumni Research Foundation and the NSF award CBET-2006028. The simulations were performed using Bridges at the Pittsburgh Supercomputing Center through allocation TG-DMR180076, which is part of the Extreme Science and Engineering Discovery Environment (XSEDE) and supported by NSF grant ACI-1548562. The work at University of Michigan is supported by NSF CAREER grant DMR-1847847.

**Author Contributions**

J.-M.H. conceived the idea, designed and supervised the research. S. Z. developed the computer codes and performed the research. Under the supervision of J.T.H., P. B. M. grew the epitaxial $Fe_{80}Ga_{20}$/MgO heterostructure and measure its magnetic parameters that are used as the input of the computational model. J.-M.H. and S.Z. wrote the paper using feedback from P.B.M. and J.T.H..

**Main Figures**

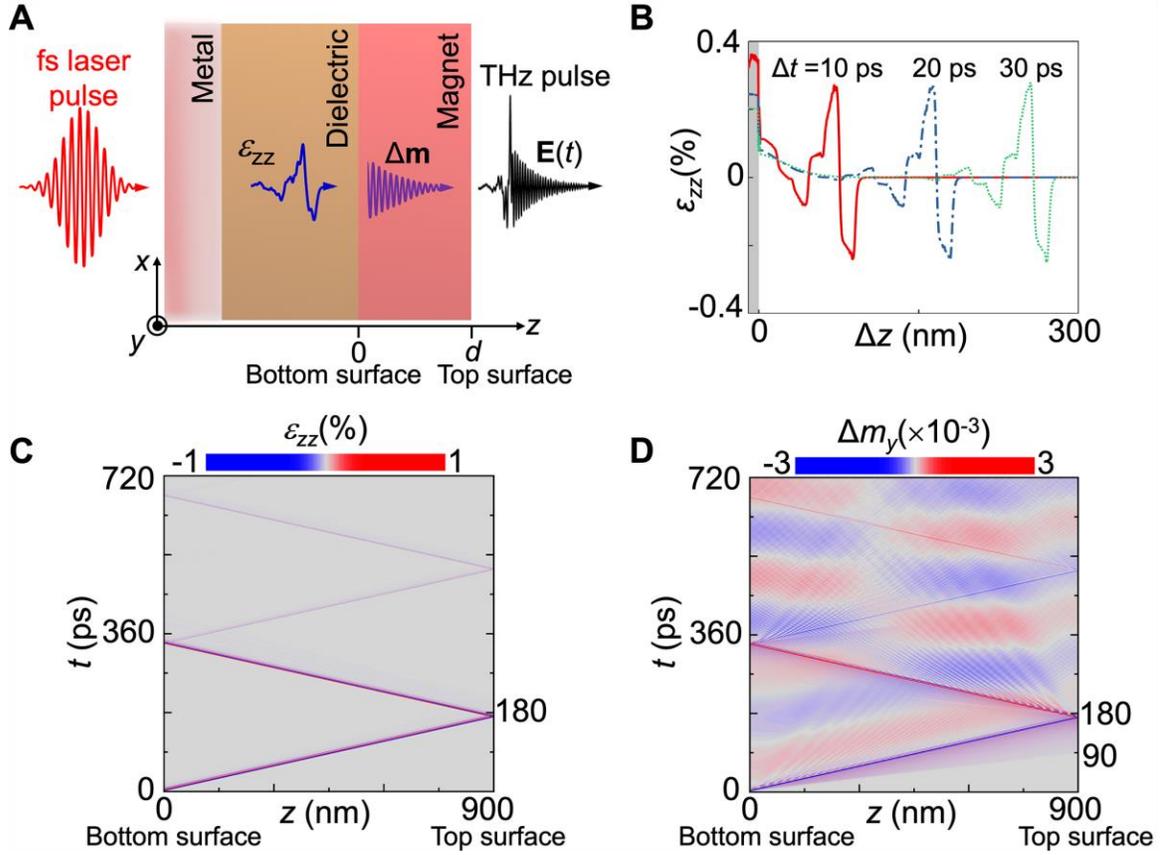

**Figure 1**. **Excitation of spin waves in magnetoelastic film by laser-induced short strain pulse**. (**A**) Schematic of the metal/dielectric/magnetoelastic heterostructure which can convert a fs laser pulse into a narrowband THz electromagnetic pulse through the excitation of spin wave $\Delta \mathbf{m}$ by a laser-induced short strain pulse $\varepsilon_{zz}$. Electric-field component $\mathbf{E}(t)$ at 2 nm above the top surface of the magnetoelastic film (the distance is exaggerated in the schematic) is calculated. (**B**) Profiles of the strain $\varepsilon_{zz}$ across the thickness direction of Al/(001)MgO at $\Delta t$ = 10 ps, 20 ps, and 30 ps after irradiating the surface of the polycrystalline Al film (indicated by shade) with a fs laser at $\Delta t$ =0 ps. Propagation of (**C**) the strain pulse $\varepsilon_{zz}(z,t)$ and (**D**) spin wave $\Delta m_y(z,t)$ between the bottom ($z$=0 nm) and top ($z$=900 nm) surfaces of the (001)Fe film. Note that $\Delta m_y = m_y(z,t) - m_y(z,t=0)$ is evaluated using the $m_y$ at $t$=0 ps as the reference, where $t$ =0 ps is defined as the moment the strain pulse propagates into the Fe film from its bottom surface.

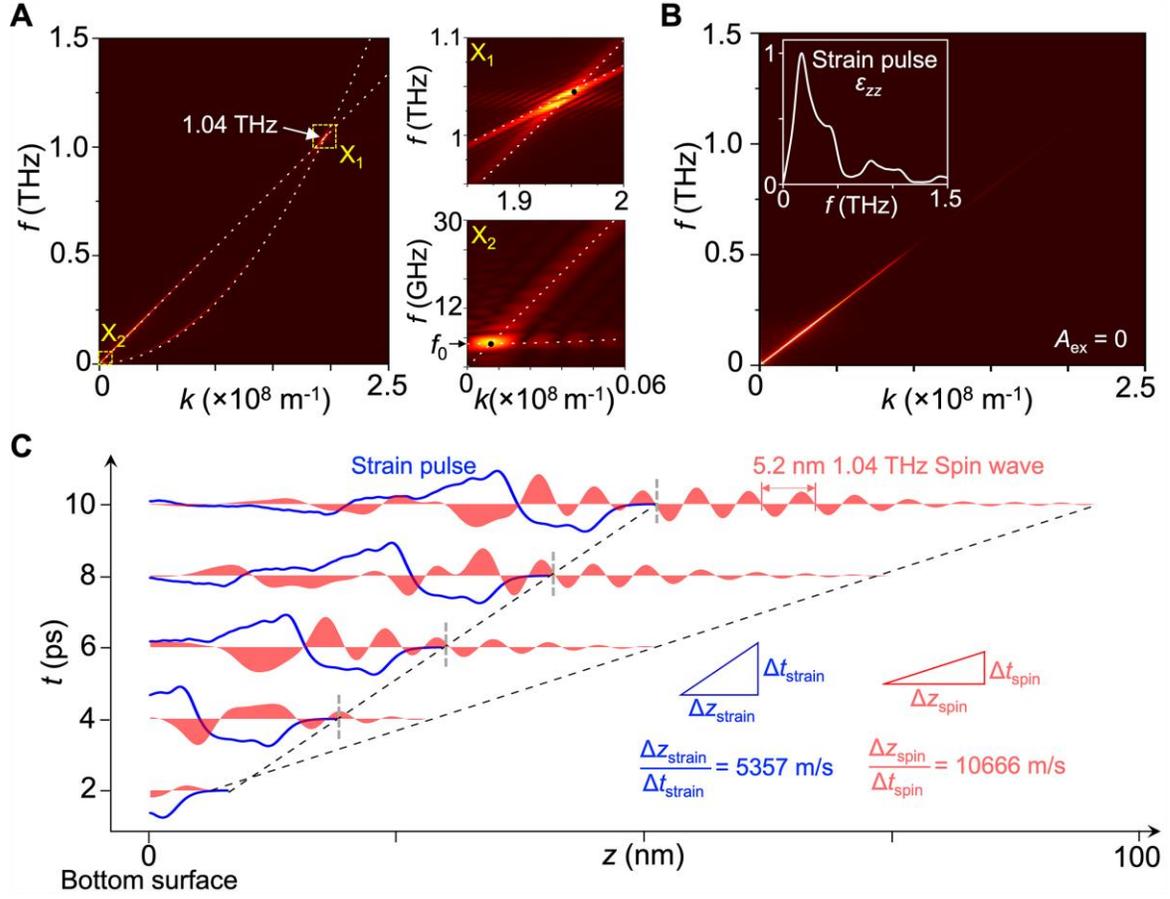

**Figure 2. Mechanisms of spin wave excitation**. Frequency-wavenumber relationship $f(k)$ of the $\Delta m_y(z,t)$ across the entire Fe film thickness from $t=0$ ps to $t=720$ ps with (**A**) $A_{ex}=21$ pJ m$^{-1}$ and (**B**) $A_{ex} = 0$. In (A), enlarged plots near the two crossing points $X_1$ and $X_2$ (indicated by black dots) are shown on the side, where $f_0$ (~4.73 GHz) in the bottom plot indicates the ferromagnetic resonance frequency. The inset in (B) shows the freqeuncy spectrum of the strain pulse in the Fe film. (**C**) Profiles of the injected strain pulse and the excited spin wave near the Fe bottom surface from $t=0$ ps to $t=10$ ps. The strain pulse is represented by blue lines and the spin wave is represented by red shades. The wave amplitudes scale with the $\varepsilon_{zz}$ and $\Delta m_z$, respectively. The frontmost positions of the strain pulse and spin waves at different moments are connected by dashed lines, and their group velocities are indicated. The waveform of the 1.04 THz spin wave, with a wavelength of 5.2 nm, starts from the frontmost position of the strain pulse (indicated by the vertical dashed line).

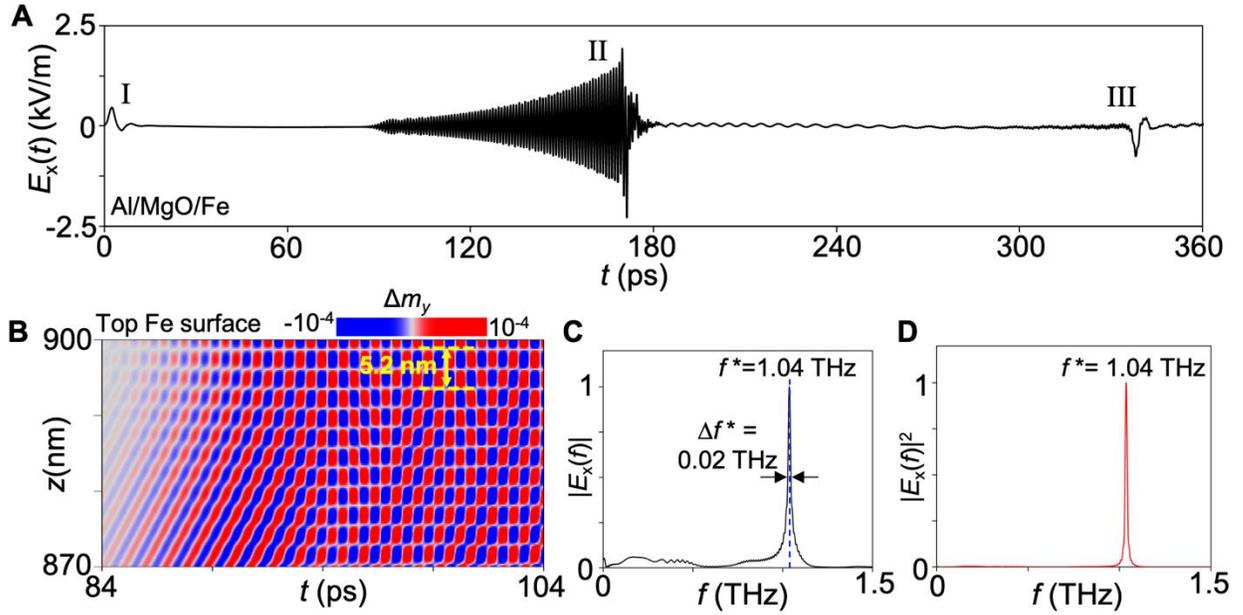

**Figure 3. THz radiation from spin waves**. (**A**) Electric-field component of the emitted electromagnetic wave $E_x(t)$, which has three distinct wave packets ($t$=0-16 ps, 84-184 ps, 252-352 ps for packet I, II, III, respectively). (**B**) Spatiotemporal profile of standing exchange spin wave $\Delta m_y(z,t)$ near the Fe top surface within $t$=84-104 ps, which has a wavelength of ~5.2 nm. (**C**) Frequency spectrum and (**D**) energy spectral density of the second wave packet of the $E_x(t)$.

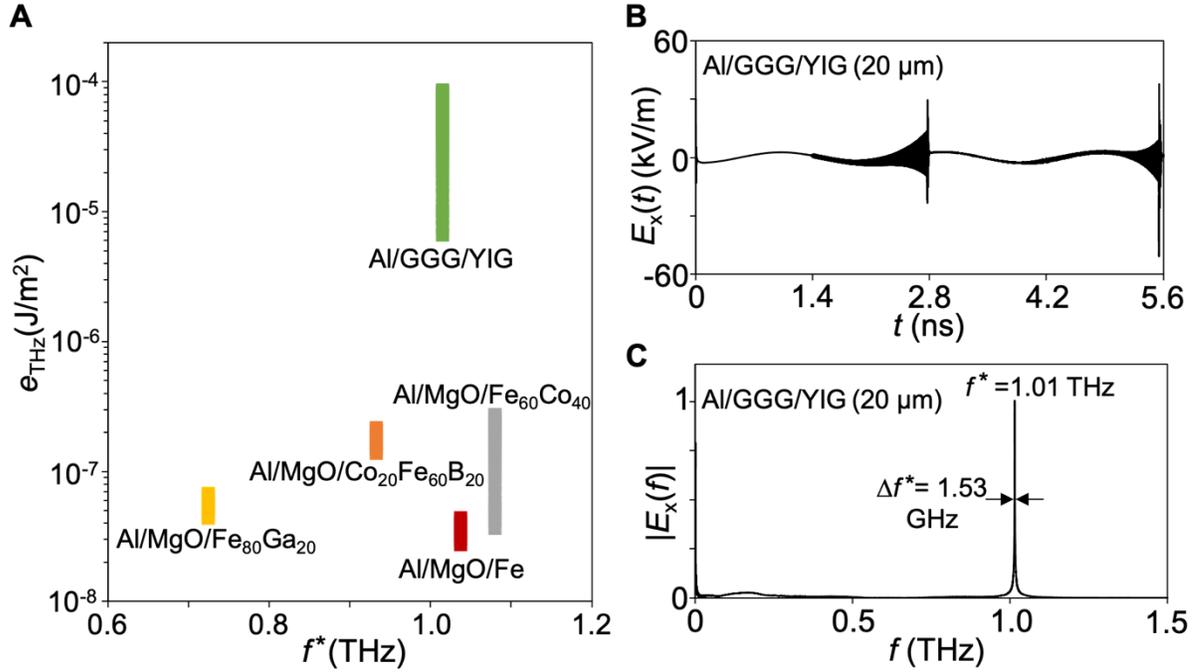

**Figure 4. Design of magnetoelastic heterostructures for tuning the energy density and frequency of the THz emission.** (**A**) Frequency $f^*$ and energy density of the monochromatic THz emission $e_{THz}$ from different metal/dielectric/magnetoelastic heterostructures, each with varying magnetoelastic layer thickness within different ranges. The thickness of the magnetoelastic layer is 250-2000 nm in Al/MgO/Fe, 200-1200 nm in Al/MgO/Fe$_{60}$Co$_{40}$, 200-1000 nm in Al/MgO/Fe$_{80}$Ga$_{20}$, 250-2000 nm in Al/MgO/Co$_{20}$Fe$_{60}$B$_{20}$, and 0.5-20 µm in Al/Gd$_3$Ga$_5$O$_{12}$(GGG)/Y$_3$Fe$_5$O$_{12}$(YIG). (**B**) The electric field $E_x(t)$ emitted from the heterostructure based on 20-µm-thick YIG, and (**C**) the frequency spectrum of its second wave packet from 1.386 ns to 2.789 ns. For accelerating the computation, the damping coefficient $\alpha$ of the YIG is set as $3\times10^{-4}$ in (B) and (C), which is larger than the reported value of $8\times10^{-5}$. Therefore, the $E_x(t)$ emitted from the YIG-based heterostructure can be larger than that the values shown in (B) in practice.

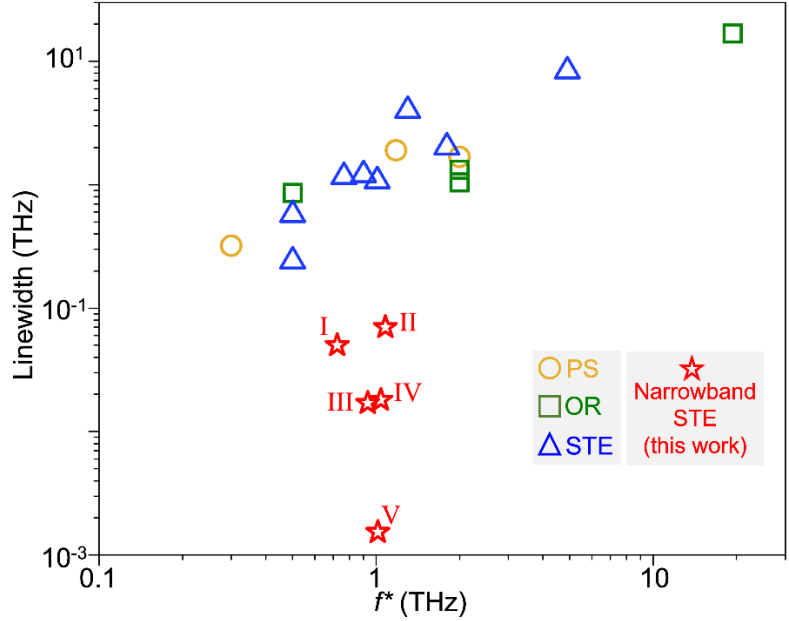

**Figure 5**. **Comparison to existing devices**. Peak frequencies $f^*$ and linewidths of the THz pulses from the present narrowband spintronic THz emitter (STE) and the three types of existing fs-laser-excited THz emitters, including photoconductive switch (PS), nonlinear crystal with optical rectification (OR) effect, and conventional STE based on metallic FM/HM stack. For the narrowband STE, the five different magnetoelastic heterostructures are Al/MgO/(001)$Fe_{80}Ga_{20}$ (I), Al/MgO/(001)$Fe_{60}Co_{40}$ (II), Al/MgO/(001)$Co_{20}Fe_{60}B_{20}$ (III), Al/MgO/(001)Fe (IV) and Al/GGG/(001)YIG (V).

For Table of Contents Only

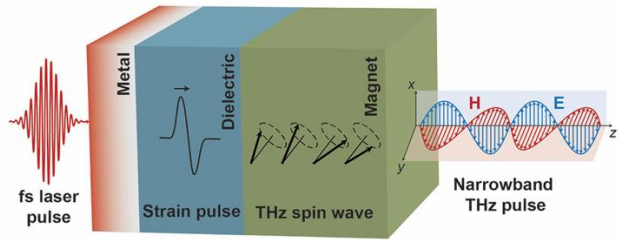